\documentclass[11pt]{article}

\usepackage{amsmath}
\usepackage{latexsym}
\usepackage{amsfonts}
\usepackage[normalem]{ulem}
\usepackage{array}
\usepackage{cite}
\usepackage{amssymb}
\usepackage{graphicx}
\usepackage{txfonts}
\usepackage{wasysym}
\usepackage{enumitem}

\usepackage{ragged2e}
\usepackage[svgnames,table]{xcolor}
\usepackage{longtable}
\usepackage{changepage}
\usepackage[none]{hyphenat}
\usepackage{setspace}
\usepackage{hhline}
\usepackage{multicol}

\usepackage{multirow}

\usepackage[toc,page]{appendix}
\usepackage[paperheight=11.0in,paperwidth=8.5in,left=0.69in,right=0.69in,top=1.0in,bottom=1.0in,headheight=1in]{geometry}
\usepackage[utf8]{inputenc}
\usepackage[T1]{fontenc}
\usepackage{hyperref}
\begin{document}

{\fontsize{20pt}{20pt}\selectfont \textbf{Rapid De Novo Shape Encoding: A Challenge to \\  Connectionist Modeling}\par}\par

{\fontsize{14pt}{14pt}\selectfont Ernest Greene\par}\par

\vspace{\baselineskip}
\textit{Laboratory for Neurometric Research, Department of Psychology, University of Southern California,\\ Los Angeles, USA}\par

\vspace{\baselineskip}

\vspace{\baselineskip}

\vspace{\baselineskip}
\begin{multicols}{2}
\begin{justify}
{\fontsize{14pt}{14pt}\selectfont Abstract\par}
\end{justify}\par

\vspace{\baselineskip}

\vspace{\baselineskip}

\end{multicols}
\begin{multicols}{2}
\begin{justify}
\textbf{Neural\ network (connectionist) models are designed to encode image features and provide the building blocks for object and shape recognition.  These models generally call for:\  a) initial diffuse connections from one neuron population to another, and  b) training to bring about a functional change in those connections so that one or more high-tier neurons will selectively respond to a specific shape stimulus.\  Advanced\ models provide for translation, size, and rotation invariance.  The present discourse notes that recent work on human perceptual skills has demonstrated immediate encoding of unknown shapes that were\ seen only once.  Further, the perceptual mechanism provided\ for translation, size, and rotation invariance.  This finding represents a challenge to connectionist models that require many training trials to achieve recognition and invariance.}

\end{justify}\par

\textit{No one can doubt that past experience is an important factor in some cases, but the attempt to explain all perception in such terms is absolutely sure to fail, for it is easy to demonstrate instances where perception is not at all influenced by past experience.} \  Wolfgang Kohler \cite{1}  

\hspace{10mm}Over\ the past century, a vast amount of experimental and theoretical effort has been devoted to discovering how our visual system registers and encodes shape information for purposes of recognizing objects.  The contributions include studies of human perception, recording neuronal responses to stimuli, evaluating brain-images of perceptual states, and computer simulation at every level of image processing.\  Notwithstanding\ the great many articles that have been published, we still do not have a clear idea of how the shape of a given object is encoded by neurons.  The lack of progress may be due to embrace of weak concepts that were derived from early experimental findings, building an edifice of theory on those concepts without consideration\ of their flaws.  The present commentary will focus on just one of the many\ issues that could be discussed.  This is the idea that encoding is accomplished by modifying connectivity among several neuron populations, with formal characterization of the process being described as neural network or connectionist modeling.\ \\
\hspace*{10mm}The connectionist models generally assume that encoding of shape information requires altering the connections among populations of neurons, or at least modifying the strength of influence, to accomplish progressively more selective responding to a stimulus pattern.\  The models commonly envision several processing stages, each being a population of neurons, with linkage\ from one stage to the next being non-specific as an initial condition.  Through exposure to examples of the shape to be encoded, there are changes in the location or strength of connections that bring about selective response to that shape from a late-stage neuron population.\  For effective encoding that models human shape-recognition skills, the functional change must accomplish several basic\ tasks:  1) It allows a given shape to be distinguished from alternative shapes, usually by activating one (or a few) neurons in the late-stage population, and with other shapes generating little or no response by those neurons.\  \  2) It provides for selective response to the candidate shape irrespective of where it was displayed on the input population, this being known as translation invariance.\  3) It provides for selective response even if the candidate shape is displayed at a different orientation, this being\ known as rotation invariance.   4) It provides for selective response even if the candidate shape is displayed at a different size, which is known as size invariance.\\
\hspace*{10mm} Fukushima \cite{2} provided a connectionist model that was especially effective for achieving some of the tasks outlined above, and its major design principles have been incorporated into a number of models that followed \cite{4,5,6,7,8}.\   His model attributes various stages of processing to specific anatomical structures of\ the mammalian visual system.  For example, the input layer is\ assumed to be modeling the photoreceptor array.  These connect to two layers of cells that respond like neurons found in primary visual cortex \cite{9,10}.\  The processing elements in these layers were designated as S-cells and C-cells, reflecting $``$simple$"$  and $``$complex$"$  selectivity of response, as described in the early neurophysiological literature \cite{9,10}.  At a beginning stage in the hierarchy, driving of\ S-cell and C-cell response is spatially locked to the location of the shape stimulus upon the photoreceptor array.  Training trials that display alternative shapes to a given retinal location bring about altered functional connectivity in the S and C layers, resulting in progressively more selective responses within the hierarchy.\  Connections from neurons having shape-specific responding then converge, through training, to provide for location invariance.\  Computer simulations found some tolerance for variation in the size of the shape to be identified, but the initial model did not build in substantial aptitude for size invariant discrimination.\\

\hspace*{10mm}The basic concepts advanced by Fukushima \cite{2} have been incorporated into a number of other connectionist models -- see \cite{11,12} for reviews.\  The VisNet model formulated by Rolls \cite{3} adopted the basic hierarchical cascade principle, differing primarily in the kind of early contour filters that were assumed and the specific rules for adaptive change in connectivity.\  In addition to providing for translation (location) invariance, it was designed to accomplish size invariance. Others have developed network connectivity principles that would also achieve viewpoint invariance \cite{5,6,7,8,9,10,11,12,13}.\\
\hspace*{10mm}The feasibility of the various connectionist models could be challenged on a number of grounds, but the key common principle at issue here is the need to provide training that alters connectivity among neurons in the hierarchy.\  Training trials are required to bring about the selectivity to specific shapes, which provides the basis for shape recognition.\  Depending on the model, the training trials also\ achieve translation, size, and/or rotation invariance.  So, an intrinsic feature of connectionist models is an initial $``$blank$"$  state wherein the shape attributes have not been encoded and will not provide responses that specify the shape until the network has been trained.\\
\hspace*{10mm}Because of the requirement for training to provide differential and selective response by network neurons, a recent set of experiments from my laboratory constitutes a significant challenge to these connectionist\ models.  Greene $\&$  Hautus \cite{14} have provided evidence for immediate encoding of shapes that have never been seen before.\  To be specific, each shape was displayed as a string of dots on an LED array, with a simultaneous brief flash of the dots providing perception of an outline boundary\ for the shape.  On a given trial the respondent would first be shown a target shape that was randomly drawn from a 450-shape\ inventory.\ \ For each target, every dot in the boundary was displayed.  A few moments later a comparison shape was flashed, this being a low-density (sparse) sampling of boundary dots from either the same shape as the target, or from a different shape.  The former was designated as a $``$matching$"$  shape,\ and the latter as $``$non-matching.$"$   All decisions were evaluated using methods develop from signal detection theory to\ provide an index of performance that was free from bias.  Judgments were found to be above chance, commonly well\ above chance, for the various task conditions that were tested, which provided clear evidence that the target shape had been encoded and the shape information was available to inform match judgments.  In other words, respondents were able to retain the shape cues provided by the target shape and recognize whether the comparison shape was providing some of those cues.\\
\hspace*{10mm}It is critical to affirm that these were $``$unknown$"$  shapes, meaning that they generally did not appear similar to\ the outlines of known objects.  Further, a\ given shape from the inventory was displayed as a target or as a non-matching comparison shape only once.  Therefore, the shape was not stored in long-term memory and there was no prior experience with a given shape that would constitute training.\  Even though this was a non-speeded task, respondents voiced their judgment about whether the two displays were $``$same$"$  or $``$different$"$  immediately after the comparison shape was shown, i.e., within a second or two.\\
\hspace*{10mm}Further, the experiments demonstrated location, size, and rotation invariance in the matching judgments.\  Clearly the visual system can encode the shape attributes and provide for shape comparisons very quickly, irrespective\ of changes in stimulus positioning, size differences, or rotation of the comparison shapes in relation to the target shapes.  The encoding process is both immediate and flexible, and any suggestion that neural connections must be modified through training is not a viable concept.\\
\hspace*{10mm}An additional point might be made that requirements to provide statistical proof of experimental effect actually serves\ to understate the challenge.  To avoid $``$ceiling effects,$"$  i.e., perfect or near perfect judgments, one must add constraints on the judgment process,\ such as displaying only sparse dots to mark the boundary.  Then the statistical evidence of treatment effect is that judgments were above chance, which seems a relatively weak basis for asking that a favorite concept be\ abandoned.  In the early parts of the 20\textsuperscript{th}\ century the work on psychophysics focused on performance by individual observers and there was substantial debate about using group data to assess perceptual mechanisms.  One might assert that it is far more important to note that without these experimental constraints on\ perception, our judgments are amazingly veridical.  Even when confronted with the strange demand to register whether two 10-microsecond displays have presented the same unknown shape or two different unknown shapes,\ an individual respondent will almost always be right if all the dots in the boundary are shown.  A simple test of this was done with one respondent, who scored at 95$\%$  correct when the pair contained the same shape, and 86$\%$  correct when the two members of the pair were different.\\
\hspace*{10mm}What, therefore, is the nature of the shape-encoding process?\ \ My intuition is that the major emphasis on crafting selective programming of neurons by modifying connectivity has been a mistake.  Clearly modification of anatomical connections or strength of connections occurs, but it does so relatively slowly and apparently requires substantial training to bring about functional outcomes.\  It more likely that shape attributes, especially the relative location of boundary markers, is converted into a message, i.e., information that is delivered over time.\  If the locations\ of\ markers can be derived by a pre-wired process, the information package that specifies those locations constitutes a summary of the shape.   That summary should be the same irrespective of the location at which the shape is displayed, which would\ account for location invariance.  Distance information that is\ gathered could be subject to normalization, which would provide size invariance.  A system that summarizes the boundary locations as a temporal code allows for shifting the starting point for the readout, which could accomplish rotation invariance.\\
\hspace*{10mm}Thinking of encoding as a\ process by which shape attributes are converted into a message differs substantially from considering it to be a structural modification of communication links within a population hierarchy.  What one wants is a fixed anatomical architecture that can quickly translate locations of stimuli into a temporal code that is malleable.\  More thought needs to be given to this approach.

\begin{justify}
\textbf{Acknowledgments.\ \  This research was funded by Quest for Truth Foundation.}
\end{justify}\par

\vspace{\baselineskip}
\vspace{\baselineskip}

\end{multicols}

\begin{thebibliography}{99}
\bibitem{1} Kohler W\  (1938)\  In: WD Ellis, A Source Book of Gestalt Psychology.\  London: Routledge $\&$  Egan Paul Ltd., London, 57.

\bibitem{2}Fukushima K\  (1980) \ \ Neocognitron: a self-organizing neural network model for a mechanism of pattern recognition unaffected by shift in position.  Biol Cybern\  36: 193-202.

\bibitem{3}Rolls ET  (1992)\ \ Neurophysiological mechanisms underlying face processing within and beyond the temporal cortical visual areas.  Phil Trans R Soc\  335: 11-21.

\bibitem{4}Wallis G,  Rolls, ET\  (1997)\ \ Invariant face and object recognition in the visual system.  Prog Neurobiol 51, 167-194.

\bibitem{5}Riesenhuber M, Poggio T\  (2000)\ \ Models of object recognition.  Nature Neurosci Suppl\textit{\  }3: 1199-1204.

\bibitem{6}Suzuki N, Hashimoto N, Kashimori\ Y, et al.  (2004)\  A\ neural model of predictive recognition in form pathway of visual cortex.  BioSystems\textit{ } 76: 33-42.

\bibitem{7}Pinto N, Cox DD, DeCarlo JJ  (2008)\ \ Why is real-world visual object recognition hard?  PLoS\textit{ }Comput Biol 4: e27. 

\bibitem{8}Rodriguez-Sanchez AJ, Tsotsos J K\  (2012)\  The roles of endstopped\ and curvature-tuned computations in a hierarchical representation of 2D shape.  PLoS One\  7: e40258.

\bibitem{9}Hubel DH, Wiesel TN\  (1959)\ \ Receptive fields of single neurons in the cat’s striate cortex.  J Physiol 148: 574-591.

\bibitem{10}Hubel DH, Wiesel TN\  (1968)\ \ Receptive fields and functional architecture of monkey striate cortex.   J Physiol\textit{\  }195: 215-243.

\bibitem{11}McClelland JL\  (2013)\  Integrating probabilistic models of perception and interactive neural networks: a historical and\ tutorial review.  Front Psychol\textit{ }4: e503.

\bibitem{12}Testolin A, Zorzi M.\  (2016)\  Probabilistic\ models and generative neural networks: towards a unified framework for modeling normal and impaired neurocognitive functions.  Front Comput Neurosci 10, e73.

\bibitem{13}Poggio T Edelman S\  (1990)\  A network that learns to recognize three-dimensional\ objects.  Nature\  343: 263-266.

\bibitem{14}Greene E,  Hautus\ MJ\  (2017)   Demonstrating invariant encoding of shapes using a matching judgment protocol.\  AIMS Neurosci\textit{ } 4: 120-147.
\end{thebibliography}
\end{document}